\documentclass[sigconf]{acmart}

\usepackage{enumitem}

\usepackage{amssymb}
\usepackage[ruled]{algorithm2e}
\usepackage{url}
\usepackage{multirow}
\usepackage{subfigure}
\usepackage{bm}
\usepackage{verbatim}
\usepackage{ragged2e}
\usepackage{caption}
\usepackage{subcaption}
\usepackage{graphicx}
\usepackage[normalem]{ulem}
\useunder{\uline}{\ul}{}
\usepackage{amsmath}
\usepackage{longtable}
\usepackage{adjustbox}
\usepackage{microtype}
\usepackage{setspace}
\usepackage{hyperref}
\usepackage{float}
\usepackage{balance}

\newcommand{\aka}

\AtBeginDocument{%
  \providecommand\BibTeX{{%
    \normalfont B\kern-0.5em{\scshape i\kern-0.25em b}\kern-0.8em\TeX}}}

\copyrightyear{2024}
\acmYear{2024}
\setcopyright{acmlicensed}
\acmConference[KDD '24] {Proceedings of the 30th ACM SIGKDD Conference on Knowledge Discovery and Data Mining }{August 25--29, 2024}{Barcelona, Spain.}
\acmBooktitle{Proceedings of the 30th ACM SIGKDD Conference on Knowledge Discovery and Data Mining (KDD '24), August 25--29, 2024, Barcelona, Spain}
\acmISBN{979-8-4007-0490-1/24/08}
\acmDOI{10.1145/3637528.3671585}

\begin{document}

\title{GradCraft: Elevating Multi-task Recommendations through Holistic Gradient Crafting}


\author{Yimeng Bai*}
\orcid{0009-0008-8874-9409}
\affiliation{
  \institution{University of Science and Technology of China}
  \city{Hefei}
  \country{China}
}
\email{baiyimeng@mail.ustc.edu.cn}
\thanks{*Work done at Kuaishou.}

\author{Yang Zhang$^{\dag}$}
\orcid{0000-0002-7863-5183}
\affiliation{
  \institution{University of Science and Technology of China}
  \city{Hefei}
  \country{China}
}
\email{zy2015@mail.ustc.edu.cn}

\author{Fuli Feng$^{\dag}$}
\orcid{0000-0002-5828-9842}
\affiliation{
  \institution{University of Science and Technology of China \& USTC Beijing Research Institute}
  \city{Hefei}
  \country{China}
}
\email{fulifeng93@gmail.com}
\thanks{$^{\dag}$Corresponding author.}

\author{Jing Lu}
\orcid{0009-0000-0718-6766}
\affiliation{%
 \institution{Kuaishou Technology}
 \city{Beijing}
 \country{China}}
\email{lvjing06@kuaishou.com}

\author{Xiaoxue Zang}
\orcid{0000-0002-5923-3429}
\affiliation{
  \institution{Kuaishou Technology}
  \city{Beijing}
  \country{China}}
\email{zangxiaoxue@kuaishou.com}

\author{Chenyi Lei}
\orcid{0000-0001-6287-3673}
\affiliation{
  \institution{Kuaishou Technology}
  \city{Beijing}
  \country{China}}
\email{leichy@mail.ustc.edu.cn}

\author{Yang Song}
\orcid{0000-0002-1714-5527}
\affiliation{
 \institution{Kuaishou Technology}
 \city{Beijing}
 \country{China}}
\email{yangsong@kuaishou.com}

\def\authors{Yimeng Bai, Yang Zhang, Fuli Feng, Jing Lu, Xiaoxue Zang, Chenyi Lei, Yang Song}

\renewcommand{\shortauthors}{Yimeng Bai et al.}

\begin{abstract}

    Recommender systems require the simultaneous optimization of multiple objectives to accurately model user interests, necessitating the application of multi-task learning methods. However, existing multi-task learning methods in recommendations overlook the specific characteristics of recommendation scenarios, falling short in achieving proper gradient balance. To address this challenge, we set the target of multi-task learning as attaining the appropriate magnitude balance and the global direction balance, and propose an innovative methodology named GradCraft in response. GradCraft dynamically adjusts gradient magnitudes to align with the maximum gradient norm, mitigating interference from gradient magnitudes for subsequent manipulation. It then employs projections to eliminate gradient conflicts in directions while considering all conflicting tasks simultaneously, theoretically guaranteeing the global resolution of direction conflicts. GradCraft ensures the concurrent achievement of appropriate magnitude balance and global direction balance, aligning with the inherent characteristics of recommendation scenarios. Both offline and online experiments attest to the efficacy of GradCraft in enhancing multi-task performance in recommendations. The source code for GradCraft can be accessed at \url{https://github.com/baiyimeng/GradCraft}.

\end{abstract}




\maketitle

\section{Introduction}\label{intro}

Recommender systems assume a pivotal role in personalized information filtering, significantly shaping individual online experiences~\cite{D2Q,RTSVR,D2Co}. The effectiveness of the systems often hinges on the ability to thoroughly model user interests, which typically entails simultaneously optimizing multiple user feedback that reflects different facets of user satisfaction~\cite{DML,LOINF,SINE}. For instance, a short video recommender system needs to optimize both the time of watching a video and the likelihood of liking it~\cite{LabelCraft,TSCAC}. Consequently, there has been an increasing trend towards applying multi-task learning in recommender systems to model the various facets of user satisfaction simultaneously~\cite{survey}, forming the mainstream approach in major industry applications~\cite{MMoE,PLE}.

Multi-task learning aims to optimize multiple objectives simultaneously. Current approaches in recommendation predominantly involve the direct application of general multi-task optimization methods from machine learning. These methods typically focus on achieving a proper balance among tasks to prevent negative transfer effects~\cite{MMoE} from two gradient perspectives. The first line of work involves reweighting loss, adjusting the gradient magnitudes based on specific criteria such as uncertainty ~\cite{Uncertainty} and update speed~\cite{DWA,GradNorm,DBMTL} to effectively balance attention across different tasks. However, these methods exhibit limitations in handling task conflicts, showing unstable performance~\cite{revisiting}, especially when confronted with significant task heterogeneity like recommendation. The second line of work concentrates on manipulating gradient directions to diminish negative cosine similarity between tasks~\cite{PCGrad,GradVac,CAGrad,MGDA,IMTL}. However, their gradient manipulation is typically executed in pairs, lacking the assurance of global non-conflict. Additionally, their manipulation overlooks interference from gradient magnitudes. These limitations significantly affect their efficacy, particularly in recommendation scenarios involving numerous tasks.

Given the pros and cons of existing methods, we distill the essence of multi-task optimization as achieving both an \textbf{appropriate magnitude balance} and a \textbf{global direction balance} on gradients, enhancing the suitability for recommendation. Firstly, it is crucial to ensure appropriate consistency in the magnitudes of gradients for heterogeneous recommendation tasks. The absence of such balance may result in certain tasks dominating others~\cite{DBMTL}, thereby leading to subpar recommendation performance. At the same time, it is also imperative to completely resolve any conflicts in gradient directions across numerous recommendation tasks concurrently, thereby ensuring global non-conflict in gradients. Failure to do so could result in residual conflicts between some tasks, hindering the transfer of knowledge and finally compromising the efficacy of multi-task optimization.

In this work, we introduce \textit{GradCraft}, a dynamic gradient balancing method for multi-task optimization. To ensure both magnitude and direction balance simultaneously, we devise a sequential paradigm that involves gradient norm alignment followed by direction projection. Initially, we dynamically align gradient norms across all tasks based on the maximum norm, establishing an appropriate magnitude balance. Subsequently, utilizing this balanced outcome, we apply projections to eliminate gradient conflicts in directions while considering all conflicting tasks concurrently, thereby ensuring global direction balance. In this sequential process, achieving direction balance hinges on attaining magnitude balance, avoiding interference from the magnitude imbalance.

Delving into further detail, in the magnitude balance, we do not pursue an absolute gradient norm alignment across different tasks; rather, we aim to prevent the norm differences from becoming too pronounced (such as spanning multiple orders of magnitude), thus averting dominance by certain tasks while preserving task specificity. In the direction balance, we move beyond mere orthogonality after projections. Our emphasis is on requiring a certain level of positive similarity to facilitate the positive transfer of knowledge across tasks, thereby enhancing conflict resolution. These design principles enable us to achieve a better balance of magnitudes and more thorough conflict resolution. We apply GradCraft to the \textit{Progressive Layered Extraction} (PLE)~\cite{PLE} model and validate it through both offline and online experiments. The resulting empirical findings consistently demonstrate GradCraft's superior performance in multi-task recommendation scenarios.

The main contributions of this work are summarized as follows:

\begin{itemize}[leftmargin=*]
    \item We underscore the significance of concurrently achieving appropriate magnitude balance and global direction balance, aligning with the characteristics inherent in recommendation scenarios.
    \item 
    We introduce GradCraft, an innovative methodology that incorporates a flexible magnitude adjustment approach followed by a global direction deconfliction strategy.
    \item 
    We systematically conduct a series of experiments, both offline and online, showcasing GradCraft's effectiveness in improving multi-task recommendations.
\end{itemize}
\section{Preliminary}
\subsection{Multi-task Recommendation}
Multi-task recommendation aims to optimize multiple recommendation objectives simultaneously. Let $\mathcal{D}$ represent the historical data. Each sample in $\mathcal{D}$ is denoted as $(\bm{x}, \bm{y})$, where $\bm{x}$ represents the features of a user-item pair, and $\bm{y}=[y_1, \dots, y_T]$ denotes $T$ distinct task labels of user behaviors, such as Effective View~\cite{DML,PEPNet} and Like. The target is to learn a multi-task recommender model $f_\theta$ that uses $\bm{x}$ to predict the labels $\bm{y}$ by fitting $\mathcal{D}$. Each task involves the prediction of a specific label $y_i$, and corresponds to a specific loss objective $\ell_i$,
which can be expressed as
\begin{equation}\label{eq:loss}
    \ell_i = L(f_\theta(\bm{x})_i, y_i; \mathcal{D}) \quad (i=1, \dots, T),
\end{equation}
where $L$ denotes the common recommendation loss function, such as Binary Cross Entropy (BCE) loss~\cite{DIL} and Mean Squared Error (MSE) loss~\cite{NCF}.
Here, for briefness, we omit the regularization term which is widely adopted to prevent overfitting.

\textit{Multi-task Optimization}. 
To optimize the multiple objectives, existing methodologies adhere to a unified paradigm: initially, the gradients of different tasks are manipulated and then combined into a single gradient using specialized methods; subsequently, the model parameters are updated according to the combined result. Each task gradient can be obtained through backpropagation. Formally, the gradient of the $i$-th task can be represented as 
\begin{equation}\label{eq:grad}
    g_i=\nabla_\theta\ell_i\in \mathbb{R}^d,
\end{equation}
where $d$ denotes the dimension of the model parameters. Here, without loss of generality,
we treat $\theta$ and its gradient as a row vector, even though their original form can be a matrix or
tensor.

\subsection{Gradient Balance}
Recommendation tasks often exhibit the significant heterogeneity across various aspects, such as data sparsity~\cite{PEPNet,TSCAC}. This heterogeneity can lead to differences in gradient magnitudes and inconsistencies in update directions among tasks, leading to potential negative transfer effects~\cite{MMoE}. To mitigate such effects, it is essential to achieve magnitude and direction balance.

\subsubsection{Magnitude Balance}
The assessment of the magnitude of a task gradient $g_i$ typically relies on its norm~\cite{MetaBalance}, denoted as $\Vert g_i \Vert$. Magnitude balance concerns the consistency in the magnitudes of different task gradients, aiming to prevent situations where tasks $i$ and $j$ exhibit a significant difference in magnitudes, expressed as
\begin{equation}\label{eq:magnitude}
    \Vert g_i \Vert \gg \Vert g_j \Vert \quad or \quad  \Vert g_i \Vert \ll \Vert g_j \Vert.
\end{equation}
The lack of magnitude balance may result in specific tasks exerting dominance over the optimization process, ultimately leading to the sub-optimal recommendation performance~\cite{DBMTL}.

\subsubsection{Direction Balance}
Direction balance is aimed at averting conflicts between different tasks, where conflicts are defined by a negative cosine similarity between two task gradients~\cite{PCGrad,GradVac}. Specifically, task gradients $g_i$ and $g_j$ are considered in conflict if the inner product between them holds
\begin{equation}\label{eq:direction}
    \langle g_i, g_j\rangle < 0.
\end{equation}
Achieving direction balance involves eliminating such negative similarities. The lack of direction balance can hinder the knowledge transfer among different recommendation tasks, finally compromising the efficacy of multi-task optimization.
\section{Methodology}
In this section, we commence by furnishing an overview of the proposed methodology. Subsequently, we introduce the magnitude adjustment approach aimed at achieving appropriate magnitude balance, and present the proposed global direction deconfliction strategy, which aims to attain global direction balance. Finally,  we delve into a discussion of our gradient projection method.

\begin{figure}[t]
    \centering
    \includegraphics[width=0.47\textwidth]{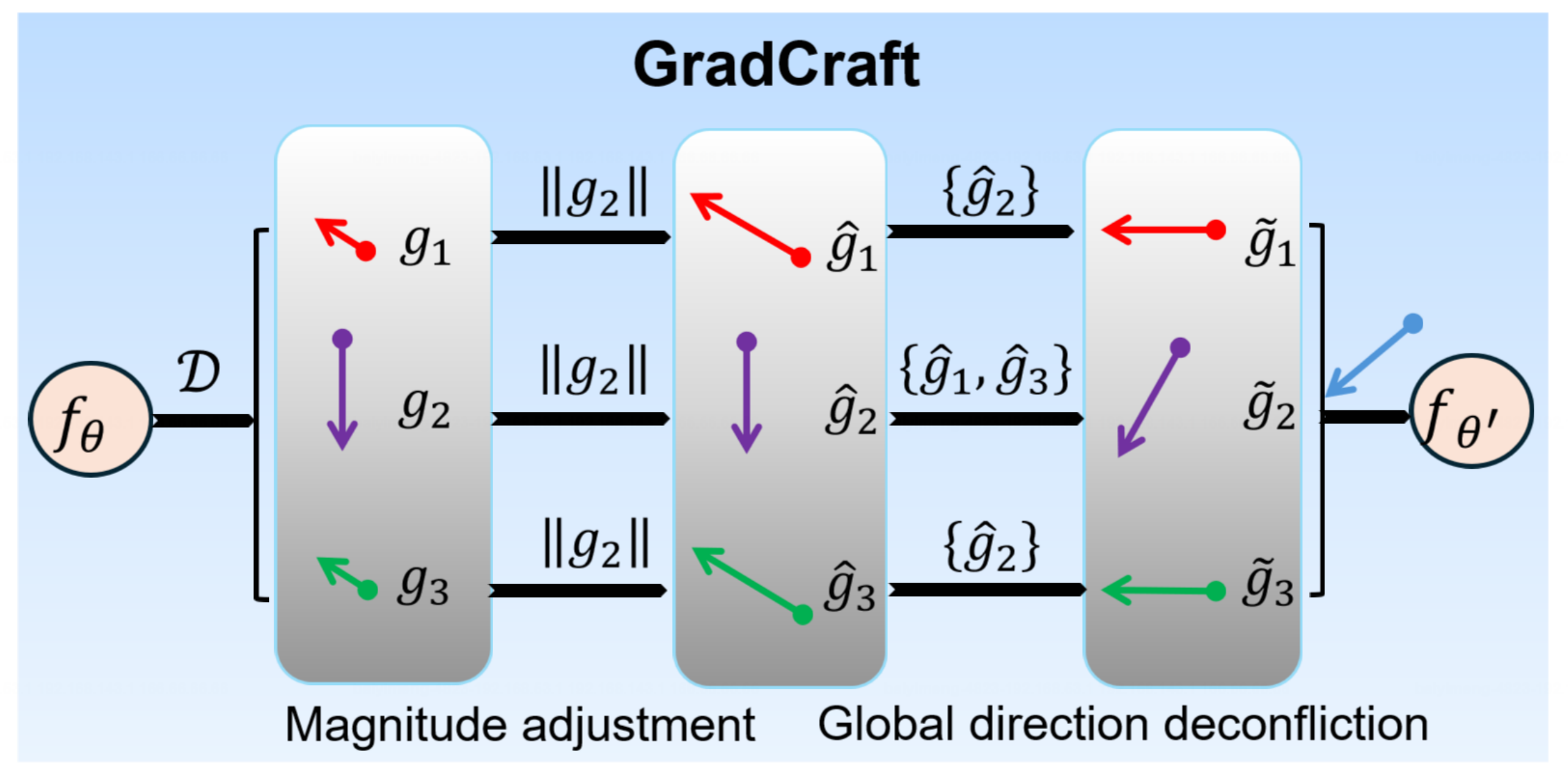}
    \caption{An overview of GradCraft. It initially adjusts the gradient magnitude based on the maximum norm. Subsequently, it performs gradient projections based on the conflicting task gradients and aggregates the gradients to update the recommender model, globally deconflicting in directions.}
    \label{fig:GradCraft}
\end{figure}

\subsection{Overview} 
We aim to achieve a simultaneous balance in both the gradient magnitude and direction. To accomplish this, we propose a sequential paradigm that involves aligning gradient norms followed by projection operations, as illustrated in Figure~\ref{fig:GradCraft}. Firstly, we dynamically align gradient norms across all tasks based on the maximum norm, establishing an appropriate balance in magnitudes. Secondly, using this balanced outcome, we apply projections to eliminate gradient conflicts while considering all conflicting tasks concurrently, ensuring a global balance in directions. Finally, we merge the gradients and update the recommender model. Given that our method operates at the gradient level, we name it \textit{GradCraft}.

\subsection{Magnitude Adjustment}
In order to mitigate interference arising from differences in gradient magnitudes across tasks, our primary focus lies in the adjustment of gradients to ensure an appropriate level of magnitude balance. Rather than pursuing absolute uniformity of gradient norms across different tasks, we aim to prevent excessive differences in norms, such as those spanning multiple orders of magnitude. This helps avert dominance by certain tasks while preserving task specificity. To achieve this, for each task, we adjust its gradient norm by combining its original norm with the maximum norm among tasks. Formally, the adjustment is performed as
\begin{equation}\label{eq:mag_adj_new}
    \hat{g}_i = \tau\frac{\max_{j}\Vert g_j\Vert}{\Vert g_i\Vert}g_i + (1-\tau)g_i,
\end{equation}
where $g_i$ represents the original gradient of task $i$, $\max_{j}\Vert g_j\Vert$ is the maximum gradient norm among all tasks, and $\hat{g}_i$ denotes the adjusted task gradient. The hyper-parameter $\tau \in [0, 1]$ undergoes tuning based on validation performance.  In this manner, we ensure that the difference between the maximum and minimum gradient norms among tasks does not exceed $\frac{1}{\tau}$ times.

\subsection{Global Direction Deconfliction}
After adjusting the magnitudes, we aim to achieve the global gradient balance. For each task, we utilize projections to ensure its gradient does not conflict with the gradients of all other tasks concurrently. Subsequently, we linearly combine the deconflicted gradients from all tasks for the final model updating.

\textbf{Gradient projection.} For a given task gradient $\hat{g}_i$, we denote the gradients conflicting with it as $G_i = [\hat{g}_{i_1}, \dots, \hat{g}_{i_n}] \in \mathbb{R}^{n\times d}$, where $\hat{g}_{i_j}$ represents the $j$-th conflicting gradient. We define a projection target to achieve non-negative similarities between the deconflicted gradient and all conflicting gradients as 
\begin{equation}\label{eq:gcp}
    \begin{split}
        G_i\tilde{g}^{\top}_i = &\bm{z}, \\
        \bm{z} = [\epsilon \Vert \hat{g}_i\Vert\Vert \hat{g}_{i_1}\Vert, \dots, & \epsilon \Vert \hat{g}_i\Vert\Vert \hat{g}_{i_n}\Vert],
    \end{split}
\end{equation}
where $\tilde{g}_i$ represents the deconflicted task gradient, and $\epsilon \ge 0$ serves as a factor for adjusting the desired similarity, with a higher value indicating higher positive similarity. Notably, instead of solely pursuing gradient orthogonality ($\epsilon =0$) between tasks, we require a certain level of positive similarity to emphasize the positive transfer of knowledge across tasks, thereby enhancing conflict resolution.

Theoretically, the desired gradient $\tilde{g}_i$ could be obtained as the sum of the original gradient and the projection onto the linear space of all conflicting gradients, which can be formulated as
\begin{equation}\label{eq:td}
    \tilde{g}_i = \hat{g}_i + \sum_{k=1}^{n} w_k \hat{g}_{i_k} = \hat{g}_i + \bm{w}^{\top}G_i,
\end{equation}
where $\bm{w}\in \mathbb{R}^{n\times1}$ is a weight vector that needs to be  determined. Combining Equation~\eqref{eq:gcp} and Equation~\eqref{eq:td}, we deduce that
\begin{equation}\label{eq:final}
    G_iG_i^{\top}\bm{w} = -G_i \hat{g}_i^{\top}+\bm{z}.
\end{equation}
Given that the dimension of model parameters significantly exceeds the number of tasks, \textit{i.e.}, $d \gg n$, it is reasonable to assume that the matrix $G_i$ possesses full rank~\cite{UIF}. Consequently, the positive definiteness of $G_iG_i^{\top}$ can be attained, enabling the weight vector $\bm{w}$ to be solved in closed form as
\begin{equation}\label{eq:weight}
    \bm{w} = (G_iG_i^{\top})^{-1}(-G_i \hat{g}_i^{\top}+\bm{z}).
\end{equation}

\textbf{Gradient combination.} 
After deconflicted gradients for all tasks are obtained, we linearly combine them and utilize the aggregated gradient to update the model, which is formulated as
\begin{equation}\label{eq:update}
    \theta^\prime = \theta - \eta \frac{1}{T} \sum_{i=1}^{T} \tilde{g}_{i},
\end{equation}
where $\eta$ denotes the learning rate.

\begin{algorithm}[t]
    \caption{GradCraft}
    \label{alg:GradCraft}
    \LinesNumbered
    \KwIn{Recommender model $f_\theta$, training dataset $\mathcal{D}$, task number $T$, hyper-parameter $\tau$ and $\epsilon$, learning rate $\eta$}
    Initialize $\theta$ randomly\;
    \While{stop condition is not reached}{
        {
        // Step 1 (computation of task gradients)\;
        \For{$i=1,\dots, T$}{
        Compute $l_i$ with Equation~\eqref{eq:loss}\;
        Compute $g_i$ with Equation~\eqref{eq:grad}\;
        }
        // Step 2 (magnitude adjustment)\;
        \For{$i=1,\dots, T$}{
        Compute $\hat{g}_i$ with Equation~\eqref{eq:mag_adj_new}\;
        }
        // Step 3 (gradient projection)\;
        \For{$i=1,\dots, T$}{
            $\tilde{g}_i=\hat{g}_i$\;  
            \If{conflicting gradient set is not empty}{
            Solve $\bm{w}$ with Equation~\eqref{eq:weight}\;
            Compute $\tilde{g}_i$ with Equation~\eqref{eq:td}\;
            }
        }
        // Step 4 (update of model parameters)\;
        Update $\theta$ with Equation~\eqref{eq:update}\;
        
        }
    }
    return $f_\theta$
\end{algorithm}

\textit{Algorithm summarization.} The intricacies of GradCraft are elucidated in Algorithm~\ref{alg:GradCraft}. During the implementation phase, updates are performed on a batch of data. In each iteration, the algorithm commences by computing all task gradients (lines 4-7). Subsequently, it applies magnitude adjustments to ensure an appropriate magnitude balance, avoiding interference from the gradient magnitude (lines 9-11). Following this, if conflicts arise among task gradients, a gradient projection method is employed to ensure a global direction balance for each task (lines 13-19). Ultimately, the gradients for different tasks are combined to update the model parameters (line 21). It is important to highlight that the update process is adaptable enough to accommodate various optimizers such as Adam~\cite{Adam} and Adagrad~\cite{Adagrad}. Besides, the update process exclusively involves updating the shared model parameters, which aligns with the approach established in previous research~\cite{PCGrad}.

\subsection{Discussion}

Our gradient projection method can be viewed as an extension of the normal projection method~\cite{PCGrad}.
Under certain circumstances, our method could degrade to approximate equality with the normal projection method. Specifically, when each given task gradient $g_i$ confronts only a single conflicting gradient denoted as $g_{i_1}$, and the hyper-parameter $\epsilon$ is set to 0, the deconflicted gradient in Equation~\eqref{eq:td} is computed as 
\begin{equation}\label{eq:pcgrad}
    \tilde{g}_i = \hat{g}_i - \frac{\langle \hat{g}_i, \hat{g}_{i_1}\rangle}{\Vert \hat{g}_{i_1}\Vert} \hat{g}_{i_1}.
\end{equation}
Disregarding the magnitude adjustment to gradients here, this computation aligns with the normal conflict projection method. 

In comparison, our method simultaneously addresses all conflicting tasks for each task while requiring a certain level of positive similarity, resulting in global and thorough conflict resolution. Notably, our method does not significantly introduce extra computation complexity. Considering $G_iG_i^{\top}\in \mathbb{R}^{n \times n}$, where $n$ is the number of conflicting task gradients for $g_i$, we can efficiently compute its inverse in Equation~\eqref{eq:weight} and obtain deconflicted gradients.
\section{Experiment}
In this section, we conduct a series of experiments to answer the following research questions:

\noindent \textbf{RQ1}: How does GradCraft perform on recommendation data compared to existing multi-task learning methods?

\noindent \textbf{RQ2}: What is the impact of the individual components of GradCraft on its effectiveness?

\noindent \textbf{RQ3}: How do the specific hyper-parameters of GradCraft influence its recommendation performance?

\noindent \textbf{RQ4}: How is the scalability of GradCraft across different levels of the gradient imbalance?

\noindent \textbf{RQ5}: How effective is GradCraft when applied to real industry recommender systems?

\subsection{Experimental Setting}
\subsubsection{Datasets}
We conduct extensive experiments on an open-world dataset and our product dataset: Wechat and Kuaishou. 
\begin{itemize}[leftmargin=*]
    \item \textbf{Wechat}. This public dataset is released as part of the WeChat Big Data Challenge\footnote{\url{https://algo.weixin.qq.com/}}, capturing user behaviors on short videos over a two-week period. To ensure dataset quality, we applied a 10-core filtering process, ensuring that each user/video has a minimum of 10 samples.
    
    \item \textbf{Kuaishou}. This dataset is sourced from our Kuaishou\footnote{\url{https://kuaishou.com/}} platform, reflecting a real-world scenario for short video recommendations. It comprises short video recommendation records for 10,000 users over a five-day period. Due to the sparser nature of the dataset, we applied a 20-core filtering process during preprocessing.
\end{itemize}

\begin{table}[t]
\caption{Statistical details of the evaluation datasets.}
\label{exp:data}
\begin{tabular}{ccccc}
\hline
Dataset  & \#User & \#Item & \#Intersection & Density \\ \hline
Wechat & 19,997 & 59,322 & 7,154,154 & 0.0060  \\
Kuaishou & 8,516 & 62,699 & 2,867,290 & 0.0054  \\ \hline
\end{tabular}
\end{table}

The summary statistics of the preprocessed datasets are presented in Table~\ref{exp:data}. Each dataset contains rich features of user and video, along with diverse user feedback. We randomly split them into training, validation, and test sets, following an 8:1:1 ratio. 

In short-video recommendation, there are two types of tasks: those related to viewing behaviors and those related to interactive behaviors. Therefore, we set user usage time and engagement as our optimization objectives, which are assessed using viewing labels and engagement labels. Specifically, we select EffectiveView (\textbf{EV})~\cite{DML}, LongView (\textbf{LV})~\cite{DML}, and CompleteView (\textbf{CV})~\cite{LabelCraft} as viewing labels. EV indicates whether the watch time of an example has exceeded 50\% of the overall watch time in the dataset, while LV indicates whether the watch time has exceeded 75\%. CV reflects whether the watch time of an example has surpassed the video duration. For engagement labels, we directly use \textbf{Like}, \textbf{Follow}, and \textbf{Forward}. All labels above are binary and fitted with BCE loss.

\subsubsection{Baselines}
We compare the proposed GradCraft with the following multi-task learning methods.
\begin{itemize}[leftmargin=*]
    \item \textbf{Single}. This approach successively assigns a weight of 1 to a specific task and assigns weights of 0 to other tasks. 
    \item \textbf{EW}. This method assigns a equal weight of $1/T$ to each task, where T represents the total number of tasks.
    \item \textbf{UC}~\cite{Uncertainty}.
    This method reweighs loss based on the uncertainty.
    \item \textbf{DWA}~\cite{DWA}. This approach adapts the loss weights by considering the update speed of the loss value.
    \item \textbf{MGDA}~\cite{MGDA}. This method manipulates gradients to achieve a local Pareto optimal solution.
    \item \textbf{PCGrad}~\cite{PCGrad}. This method addresses the gradient conflict in directions by the pair-wise projection.
    \item \textbf{GradVac}~\cite{GradVac}. This approach sets adaptive gradient similarity objectives in a learnable manner to improve PCGrad.
    \item \textbf{CAGrad}~\cite{CAGrad}. This method identifies the optimal update vector within a ball around the average gradient, maximizing the worst local improvement between tasks.
    \item \textbf{IMTL}~\cite{IMTL}. This approach learns weights to ensure that the aggregated gradient has equal projections onto each task gradient.
    \item \textbf{DBMTL}~\cite{DBMTL}. This approach guarantees that all task gradients share the same magnitude as the maximum gradient norm.
\end{itemize}
As our gradient projection method can be considered an extension of the normal projection method in PCGrad, we also introduce a variant of PCGrad to ensure fair comparisons, denoted as
\begin{itemize}[leftmargin=*]
    \item \textbf{PCGrad+}. This variant takes into account magnitude balance and adjusts gradient magnitudes based on Equation~\eqref{eq:mag_adj_new}, building upon the foundation of PCGrad.
\end{itemize}

\subsubsection{Evaluation Metrics}
In order to conduct a comprehensive evaluation of performance with respect to optimizing multiple recommendation objectives, we employ two widely recognized accuracy metrics: AUC and GAUC~\cite{PEPNet}. Following previous work~\cite{random, dense_survey, DBMTL}, we mainly focus on the average performance across all tasks. Specifically, we utilize both the average metric across all tasks and the relative metric improvement compared with the Single baseline across all tasks, which can be expressed as
\begin{equation}
    \text{AV-A}(\mathcal{M}) = \frac{1}{T}\sum_{i=1}^{T}\text{AUC}_i(\mathcal{M}),  
\end{equation}
\begin{equation}
    \text{AV-G}(\mathcal{M}) = \frac{1}{T}\sum_{i=1}^{T}\text{GAUC}_i(\mathcal{M}),
\end{equation}
\begin{equation}
    \text{RI-A}(\mathcal{M}) =  \frac{1}{T}\sum_{i=1}^{T}\frac{\text{AUC}_i(\mathcal{M})-\text{AUC}_i(Single)}{\text{AUC}_i(Single)},
\end{equation}
\begin{equation}
    \text{RI-G}(\mathcal{M}) =  \frac{1}{T}\sum_{i=1}^{T}\frac{\text{GAUC}_i(\mathcal{M})-\text{GAUC}_i(Single)}{\text{GAUC}_i(Single)}.
\end{equation}
Here, $\mathcal{M}$ represents the specific multi-task learning method, with AV-A and AV-G denoting the average value of AUC and GAUC, respectively. Similarly, RI-A and RI-G signify the relative improvement in AUC and GAUC, respectively. Across all metrics, higher values indicate better recommendation results.

\subsubsection{Implementation Details}
To ensure fair comparisons, we employ the PLE~\cite{PLE} model as the backbone recommender model for all the methods under consideration. Each task is composed of a shared expert, a task-specific expert, a gate network, and a tower network. The experts are instantiated as DeepFM~\cite{DeepFM}, combining a Factorization Machine (FM)~\cite{FM} component with a Multi-Layer Perceptron (MLP)~\cite{SMLP4Rec} module. The hidden layer configuration for the MLP is set to $256\times128\times64$. The tower network is implemented as an MLP with a hidden layer configuration of $32\times16$. The gate network structure is based on a linear layer with Softmax~\cite{SSM} serving as the activation function. The embedding size is consistently set to 16 for all user and video features.

In terms of model optimization, we employ the Adam optimizer~\cite{Adam}, setting the maximum number of optimization epochs to 1000. Optimal models are identified based on validation results, utilizing an early stopping strategy with a patience setting of 10. Parameters for the backbone recommender model are initialized using a Gaussian distribution, where the mean is fixed at 0, and the standard deviation is set to $0.01$. The dropout ratio is set to 0.2. We leverage the grid search to find the best hyper-parameters. For our method and all baselines, we search the learning rate in the range of $\{$1$e$-4, 5$e$-4, 1$e$-3$\}$, the size of mini-batch in the range of $\{$2048, 4096$\}$, and the $L_2$ regularization coefficient in $\{$0, 1$e$-6, 1$e$-5, 1$e$-4, 1$e$-3$\}$. For the special hyper-parameters of baselines, we search most of them in the ranges provided by their papers.  
Regarding our methodology, the hyper-parameter $\tau$ in Equation~\eqref{eq:mag_adj_new} to regulate the closeness to the maximum norm is searched within the interval $[0, 1]$ using a step size of 0.1, and the hyper-parameter $\epsilon$ in Equation~\eqref{eq:gcp} to achieve the desired similarity is searched in the range of $\{$0, 1$e$-12, 1$e$-11, 1$e$-10, 1$e$-9, 1$e$-8, 1$e$-7$\}$.

\subsection{Performance Comparison (RQ1)}

\begin{table*}[t]
\caption{Performance comparison between the baselines and our GradCraft on Wechat and Kuaishou, where the best results are highlighted in bold and sub-optimal results are underlined. The labels Follow and Forward are respectively abbreviated as Fol and For for simplicity. AV-A and AV-G denote the average value of AUC and GAUC across different tasks, respectively. Similarly, RI-A and RI-G signify the relative improvements of AUC and GAUC.}
\label{exp:main}

\scalebox{0.96}
{
\begin{tabular}{cccccccccccccc}
\hline
\multicolumn{14}{c}{Wechat}                                                                                                                                                                                                                             \\
\multicolumn{2}{c|}{Method}                       & Single       & EW              & UC           & DWA             & MGDA   & PCGrad          & PCGrad+      & GradVac         & CAGrad & IMTL         & \multicolumn{1}{c|}{DBMTL}  & GradCraft       \\ \hline
\multirow{8}{*}{AUC}  & \multicolumn{1}{c|}{EV}   & 0.7641       & 0.7641          & 0.7633       & 0.7646          & 0.7569 & {\ul 0.7651}    & 0.7644       & 0.7648          & 0.7647 & 0.7629       & \multicolumn{1}{c|}{0.7636} & \textbf{0.7653} \\
                      & \multicolumn{1}{c|}{LV}   & 0.8484       & 0.8484          & 0.8479       & {\ul 0.8490}    & 0.8429 & \textbf{0.8491} & 0.8486       & 0.8489          & 0.8489 & 0.8478       & \multicolumn{1}{c|}{0.8479} & {\ul 0.8490}    \\
                      & \multicolumn{1}{c|}{CV}   & 0.7610       & 0.7604          & 0.7596       & \textbf{0.7620} & 0.7515 & 0.7614          & 0.7611       & 0.7613          & 0.7614 & 0.7589       & \multicolumn{1}{c|}{0.7597} & {\ul 0.7616}    \\
                      & \multicolumn{1}{c|}{Like} & 0.8661       & 0.8664          & {\ul 0.8671} & 0.8656          & 0.8604 & \textbf{0.8675} & 0.8668       & 0.8665          & 0.8662 & 0.8669       & \multicolumn{1}{c|}{0.8650} & 0.8661          \\
                      & \multicolumn{1}{c|}{Fol}  & {\ul 0.8829} & 0.8810          & 0.8763       & 0.8809          & 0.8803 & 0.8825          & 0.8827       & 0.8791          & 0.8801 & 0.8827       & \multicolumn{1}{c|}{0.8750} & \textbf{0.8888} \\
                      & \multicolumn{1}{c|}{For}  & 0.8940       & \textbf{0.9012} & 0.9006       & 0.8983          & 0.8937 & 0.8968          & 0.9000       & 0.8991          & 0.9003 & {\ul 0.9008} & \multicolumn{1}{c|}{0.8987} & 0.9001          \\ \cline{2-14} 
                      & \multicolumn{1}{c|}{\textbf{AV-A}}   & 0.8361       & 0.8369          & 0.8358       & 0.8367          & 0.8309 & 0.8371          & {\ul 0.8373} & 0.8366          & 0.8369 & 0.8367       & \multicolumn{1}{c|}{0.8350} & \textbf{0.8385} \\
                      & \multicolumn{1}{c|}{\textbf{RI-A}}  & 0.000\%        & 0.091\%           & -0.038\%       & 0.078\%           & -0.639\% & 0.118\%           & {\ul 0.135\%}  & 0.065\%           & 0.099\%  & 0.056\%        & \multicolumn{1}{c|}{-0.129\%} & \textbf{0.278\%}  \\ \hline
\multirow{8}{*}{GAUC} & \multicolumn{1}{c|}{EV}   & 0.6207       & 0.6209          & 0.6194       & 0.6189          & 0.6055 & \textbf{0.6226} & 0.6195       & 0.6218          & 0.6200 & 0.6201       & \multicolumn{1}{c|}{0.6178} & {\ul 0.6221}    \\
                      & \multicolumn{1}{c|}{LV}   & 0.7731       & 0.7745          & 0.7740       & 0.7739          & 0.7684 & {\ul 0.7754}    & 0.7736       & \textbf{0.7755} & 0.7743 & 0.7742       & \multicolumn{1}{c|}{0.7732} & 0.7751          \\
                      & \multicolumn{1}{c|}{CV}   & 0.6499       & 0.6503          & 0.6489       & 0.6499          & 0.6345 & {\ul 0.6515}    & 0.6493       & 0.6509          & 0.6491 & 0.6488       & \multicolumn{1}{c|}{0.6464} & \textbf{0.6518} \\
                      & \multicolumn{1}{c|}{Like} & 0.6324       & 0.6382          & {\ul 0.6405} & 0.6368          & 0.6328 & \textbf{0.6422} & 0.6380       & 0.6384          & 0.6390 & 0.6393       & \multicolumn{1}{c|}{0.6385} & 0.6383          \\
                      & \multicolumn{1}{c|}{Fol}  & 0.6847       & 0.6820          & {\ul 0.6962} & 0.6915          & 0.6874 & 0.6899          & 0.6870       & 0.6721          & 0.6930 & 0.6894       & \multicolumn{1}{c|}{0.6896} & \textbf{0.7003} \\
                      & \multicolumn{1}{c|}{For}  & 0.7012       & 0.7129          & 0.7154       & 0.7141          & 0.7021 & {\ul 0.7164}    & 0.7140       & 0.7152          & 0.7135 & 0.7144       & \multicolumn{1}{c|}{0.7124} & \textbf{0.7176} \\ \cline{2-14} 
                      & \multicolumn{1}{c|}{\textbf{AV-G}}   & 0.6770       & 0.6798          & 0.6824       & 0.6809          & 0.6718 & {\ul 0.6830}    & 0.6802       & 0.6790          & 0.6815 & 0.6810       & \multicolumn{1}{c|}{0.6796} & \textbf{0.6842} \\
                      & \multicolumn{1}{c|}{\textbf{RI-G}}  & 0.000\%        & 0.413\%           & 0.791\%        & 0.559\%           & -0.809\% & {\ul 0.887\%}     & 0.472\%        & 0.288\%           & 0.653\%  & 0.589\%        & \multicolumn{1}{c|}{0.380\%}  & \textbf{1.056\%}  \\ \hline
\end{tabular}
}
\vspace{+0pt}

\scalebox{0.96}
{
\begin{tabular}{cccccccccccccc}
\hline
\multicolumn{14}{c}{Kuaishou}                                                                                                                                                                                                                           \\
\multicolumn{2}{c|}{Method}                       & Single & EW           & UC              & DWA          & MGDA   & PCGrad          & PCGrad+      & GradVac      & CAGrad & IMTL            & \multicolumn{1}{c|}{DBMTL}           & GradCraft       \\ \hline
\multirow{8}{*}{AUC}  & \multicolumn{1}{c|}{EV}   & 0.7569 & {\ul 0.7581} & \textbf{0.7582} & 0.7575       & 0.7400 & 0.7558          & 0.7564       & 0.7556       & 0.7560 & 0.7579          & \multicolumn{1}{c|}{0.7568}          & 0.7565          \\
                      & \multicolumn{1}{c|}{LV}   & 0.8263 & 0.8269       & \textbf{0.8275} & 0.8266       & 0.8143 & 0.8263          & 0.8265       & 0.8264       & 0.8266 & {\ul 0.8273}    & \multicolumn{1}{c|}{0.8265}          & 0.8265          \\
                      & \multicolumn{1}{c|}{CV}   & 0.8550 & 0.8559       & \textbf{0.8561} & 0.8555       & 0.8421 & 0.8551          & 0.8548       & 0.8551       & 0.8548 & {\ul 0.8560}    & \multicolumn{1}{c|}{0.8547}          & 0.8548          \\
                      & \multicolumn{1}{c|}{Like} & 0.9347 & 0.9287       & 0.9310          & 0.9303       & 0.9297 & 0.9325          & {\ul 0.9345} & 0.9329       & 0.9340 & 0.9307          & \multicolumn{1}{c|}{0.9343}          & \textbf{0.9347} \\
                      & \multicolumn{1}{c|}{Fol}  & 0.8322 & 0.8463       & 0.8503          & 0.8469       & 0.8430 & 0.8444          & {\ul 0.8586} & 0.8437       & 0.8581 & 0.8503          & \multicolumn{1}{c|}{0.8555}          & \textbf{0.8592} \\
                      & \multicolumn{1}{c|}{For}  & 0.8156 & 0.8180       & 0.8163          & 0.8133       & 0.8118 & 0.8241          & {\ul 0.8302} & 0.8239       & 0.8288 & 0.8171          & \multicolumn{1}{c|}{0.8267}          & \textbf{0.8309} \\ \cline{2-14} 
                      & \multicolumn{1}{c|}{\textbf{AV-A}}   & 0.8368 & 0.8390       & 0.8399          & 0.8384       & 0.8302 & 0.8397          & {\ul 0.8435} & 0.8396       & 0.8431 & 0.8399          & \multicolumn{1}{c|}{0.8424}          & \textbf{0.8438} \\
                      & \multicolumn{1}{c|}{\textbf{RI-A}}  & 0.000\%  & 0.280\%        & 0.383\%           & 0.197\%        & -0.817\% & 0.355\%           & {\ul 0.811\%}  & 0.342\%        & 0.758\%  & 0.379\%           & \multicolumn{1}{c|}{0.682\%}           & \textbf{0.844\%}  \\ \hline
\multirow{8}{*}{GAUC} & \multicolumn{1}{c|}{EV}   & 0.6724 & {\ul 0.6746} & \textbf{0.6749} & 0.6738       & 0.6546 & 0.6721          & 0.6715       & 0.6718       & 0.6719 & 0.6742          & \multicolumn{1}{c|}{0.6730}          & 0.6718          \\
                      & \multicolumn{1}{c|}{LV}   & 0.7798 & 0.7800       & \textbf{0.7810} & 0.7797       & 0.7689 & 0.7797          & {\ul 0.7802} & 0.7794       & 0.7800 & 0.7801          & \multicolumn{1}{c|}{0.7799}          & 0.7801          \\
                      & \multicolumn{1}{c|}{CV}   & 0.8317 & 0.8317       & \textbf{0.8326} & 0.8313       & 0.8223 & 0.8316          & 0.8315       & 0.8317       & 0.8316 & {\ul 0.8321}    & \multicolumn{1}{c|}{0.8314}          & 0.8315          \\
                      & \multicolumn{1}{c|}{Like} & 0.6556 & 0.6617       & 0.6621          & 0.6616       & 0.6417 & \textbf{0.6661} & 0.6605       & {\ul 0.6647} & 0.6574 & 0.6624          & \multicolumn{1}{c|}{0.6621}          & 0.6600          \\
                      & \multicolumn{1}{c|}{Fol}  & 0.5987 & 0.6443       & 0.6529          & {\ul 0.6603} & 0.6176 & 0.6349          & 0.6525       & 0.6297       & 0.6490 & \textbf{0.6629} & \multicolumn{1}{c|}{0.6375}          & 0.6567          \\
                      & \multicolumn{1}{c|}{For}  & 0.5714 & 0.6318       & 0.6287          & 0.6299       & 0.6108 & 0.6393          & 0.6422       & 0.6405       & 0.6370 & 0.6253          & \multicolumn{1}{c|}{\textbf{0.6450}} & {\ul 0.6425}    \\ \cline{2-14} 
                      & \multicolumn{1}{c|}{\textbf{AV-G}}   & 0.6849 & 0.7040       & 0.7054          & 0.7061       & 0.6860 & 0.7039          & {\ul 0.7064} & 0.7030       & 0.7045 & 0.7062          & \multicolumn{1}{c|}{0.7048}          & \textbf{0.7071} \\
                      & \multicolumn{1}{c|}{\textbf{RI-G}}  & 0.000\%  & 3.248\%        & 3.451\%           & 3.601\%        & 0.457\%  & 3.243\%           & {\ul 3.671\%}  & 3.087\%        & 3.351\%  & 3.594\%           & \multicolumn{1}{c|}{3.402\%}           & \textbf{3.791\%}  \\ \hline
\end{tabular}
}

\end{table*}

We begin by assessing the overall performance of the compared methods in optimizing multiple objectives. The summarized results are presented in Table~\ref{exp:main}, yielding the following observations:

\begin{itemize}[leftmargin=*]
    \item GradCraft demonstrates superior performance compared to the baselines on both datasets, excelling in metrics of AV-A, AV-G, RI-A, and RI-G. This highlights its ability to achieve the appropriate magnitude balance and global direction balance, showcasing its efficacy in multi-task optimization.

    \item Although PCGrad+ shows improvement over PCGrad by integrating the magnitude balance, it still falls short of GradCraft. This suggests that GradCraft's global gradient projection method surpasses PCGrad's pair-wise projection method, leading to the global and thorough direction deconfliction.
    
    \item In contrast, loss reweighting methods such as EW, UC, and DWA exhibit poor performance. Their reliance on overall loss values, without granular gradient analysis, limits their effectiveness in enhancing multi-task optimization. This highlights the importance of taking into account more fine-grained gradient magnitude and direction for enhanced performance.
    
    \item Methods that exclusively prioritize either magnitude balance or direction balance struggle to achieve optimal recommendation performance and may even lead to degradation (MGDA). This emphasizes the need of holistically addressing both magnitude and direction balance in multi-task recommendations.
    
\end{itemize}

\subsection{Ablation Study (RQ2)}

To enhance the multi-task recommendation performance in GradCraft, we propose the incorporation of a magnitude adjustment approach and a gradient projection method, with two hyper-parameters $\tau$ and $\epsilon$. To substantiate the rationale behind these design decisions, we conduct an exhaustive evaluation by systematically disabling one critical design element at a time to obtain various variants. Specifically, the following variants are introduced:

\begin{itemize}[leftmargin=*]
    \item \textbf{GradCraft-fix $\bm{\epsilon}$}, which sets $\epsilon$ to 0 and uses zero vector as the projection target in Equation~\eqref{eq:gcp};
    \item \textbf{GradCraft-fix $\bm{\tau}$}, which sets $\tau$ to 1 and disables the control of the proximity of task gradients in Equation~\eqref{eq:mag_adj_new};
    \item \textbf{GradCraft-ori}, which removes the magnitude adjustment and preserves the original magnitudes without any alteration;
    \item \textbf{GradCraft-local}, which removes the global gradient projection and replaces it with the normal projection in PCGrad.
\end{itemize}

Table~\ref{exp:abl} illustrates the comparison results on Wechat, from which we draw the following observations:

\begin{itemize}[leftmargin=*]

    \item When GradCraft disables the factors ${\epsilon}$ and the ${\tau}$, there are decreases in performance across all metrics. These results confirm the pivotal role of $\epsilon$ in maintaining a certain level of positive similarity to facilitate the transfer of knowledge across tasks, and $\tau$ in controlling magnitude proximity levels.

    \item Comparatively, GradCraft-ori outperforms GradCraft-fix $\tau$. These variants correspond to aligning the magnitude with the maximum norm and retaining the original magnitude, with $\tau$ set to 1 and 0, respectively. This observation suggests that indiscriminate adjustment of magnitude to match the maximum norm may detrimentally impact recommendation performance, underscoring the significance of appropriate proximity.

    \item The performance of GradCraft-ori and GradCraft-local is similar, indicating no advantage of global gradient projection over the normal projection strategy when magnitude adjustment is absent. However, the performance gap between GradCraft and PCGrad+ in Table~\ref{exp:main} underscores the superiority of the gradient projection method. This outcome can be attributed to disruption caused by magnitudes, underscoring the critical role of initially adjusting magnitudes to achieve magnitude balance.

\end{itemize}

\begin{table}[t]
\caption{Results of the ablation study for our GradCraft method on Wechat.}
\label{exp:abl}
\begin{tabular}{ccccc}
\hline
Method               & AV-A                   & RI-A                  & AV-G                   & RI-G                  \\ \hline
GradCraft            & 0.8385               & 0.278\%                & 0.6842               & 1.056\%                \\
GradCraft-fix ${\epsilon}$                  & 0.8382               & 0.250\%                & 0.6837               & 0.981\%                \\
GradCraft-fix ${\tau}$                 & 0.8365               & 0.039\%                & 0.6798               & 0.392\%                \\
GradCraft-ori                 & 0.8370               & 0.113\%                & 0.6835               & 0.959\%                \\
GradCraft-local                 & 0.8371               & 0.118\%                & 0.6830               & 0.887\%                \\ \hline
\multicolumn{1}{l}{} & \multicolumn{1}{l}{} & \multicolumn{1}{l}{} & \multicolumn{1}{l}{} & \multicolumn{1}{l}{} \\
\multicolumn{1}{l}{} & \multicolumn{1}{l}{} & \multicolumn{1}{l}{} & \multicolumn{1}{l}{} & \multicolumn{1}{l}{} \\
\multicolumn{1}{l}{} & \multicolumn{1}{l}{} & \multicolumn{1}{l}{} & \multicolumn{1}{l}{} & \multicolumn{1}{l}{} \\
\multicolumn{1}{l}{} & \multicolumn{1}{l}{} & \multicolumn{1}{l}{} & \multicolumn{1}{l}{} & \multicolumn{1}{l}{}
\end{tabular}
\vspace{-40pt}
\end{table}

\subsection{In-depth Analysis (RQ3 \& RQ4)}

\begin{figure}
    \centering
    \includegraphics[width=0.47\textwidth]{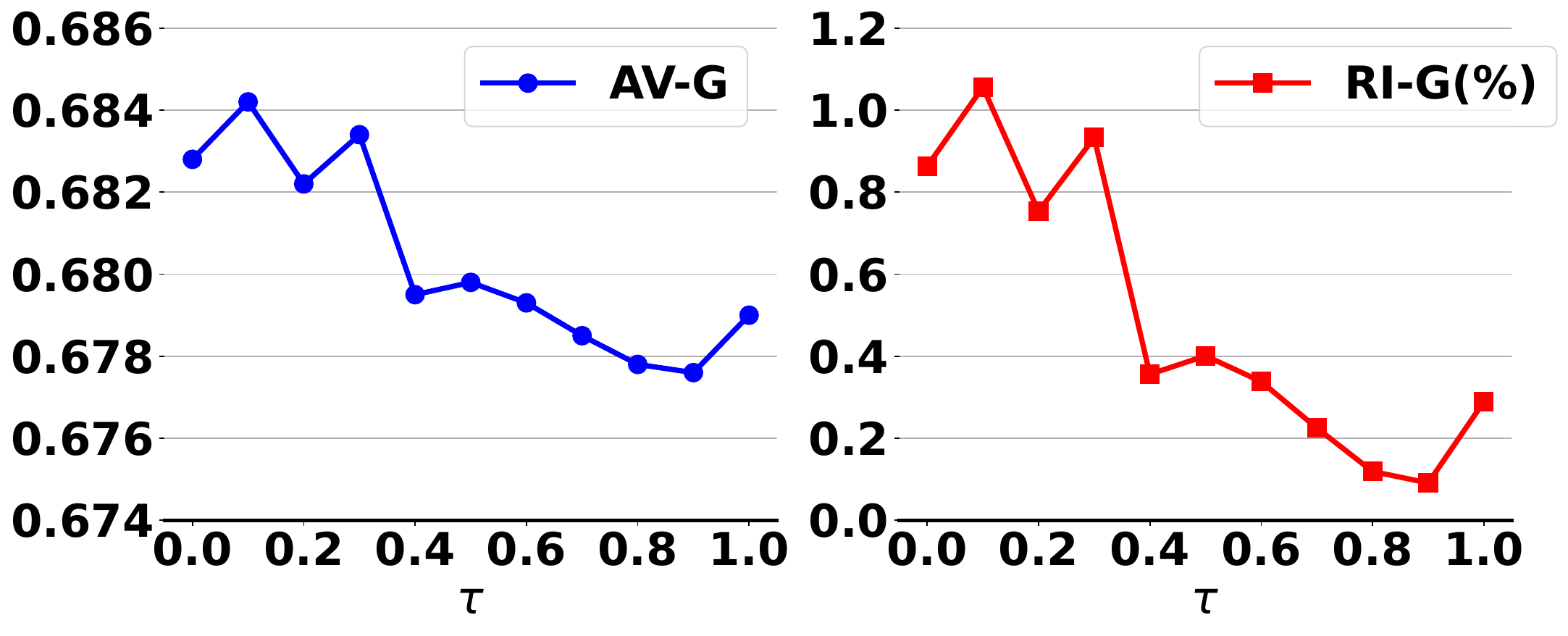}
    \caption{Results of the performance of GradCraft across different values of $\tau$ on Wechat.}
    \label{fig:hyper_tau}
\end{figure}

\begin{figure}
    \centering
    \includegraphics[width=0.47\textwidth]{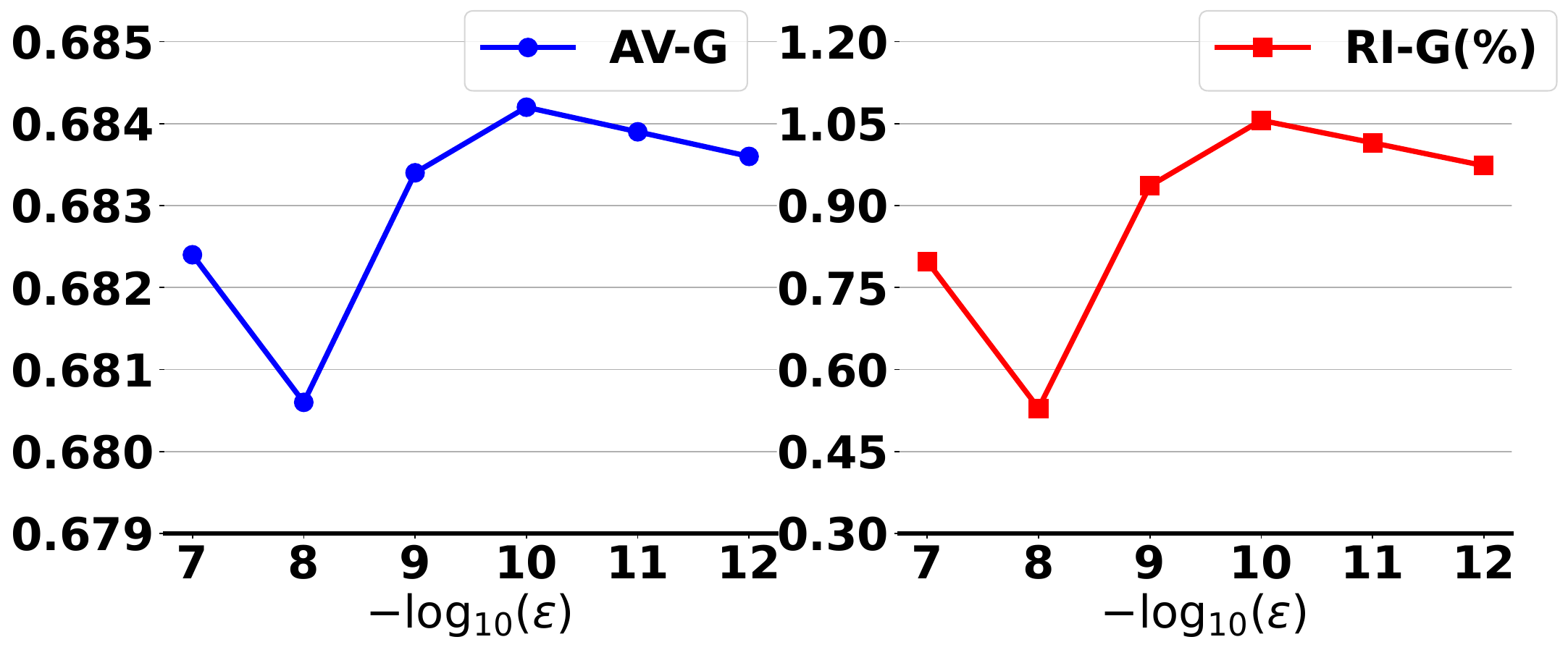}
    \caption{Results of the performance of GradCraft across different values of $\epsilon$ on Wechat.}
    \label{fig:hyper_eps}
\end{figure}

\subsubsection{The Effect of Hyper-parameter $\tau$ \& $\epsilon$}

In our investigation, the two factors $\tau$ and $\epsilon$ assume pivotal roles in influencing the effectiveness of GradCraft. We undertake a systematic examination to scrutinize the impact of varying them on the performance. 
We report the AV-G and RI-G for simplicity, as shown in Figure~\ref{fig:hyper_tau} and Figure~\ref{fig:hyper_eps}. It becomes evident that GradCraft achieves optimal AV-G and RI-G when $\tau$ is set to 0.1 and $\epsilon$ is set to 1$e$-10. However, the performance tends to deteriorate when they become excessively large. This underscores the significance of selecting an appropriate value for $\tau$ and $\epsilon$. Further analysis reveals that when $\tau\in[0, 0.3]$ and $\epsilon\in[1e-12, 1e-9]$, the performance remains consistently stable, indicating the robustness within the range. This stability is crucial for ensuring reliable performance of the magnitude adjustment approach and the gradient projection method. 

\subsubsection{The Effect of Task Number $T$}

\begin{figure}[t]
    \centering
    \includegraphics[width=0.47\textwidth]{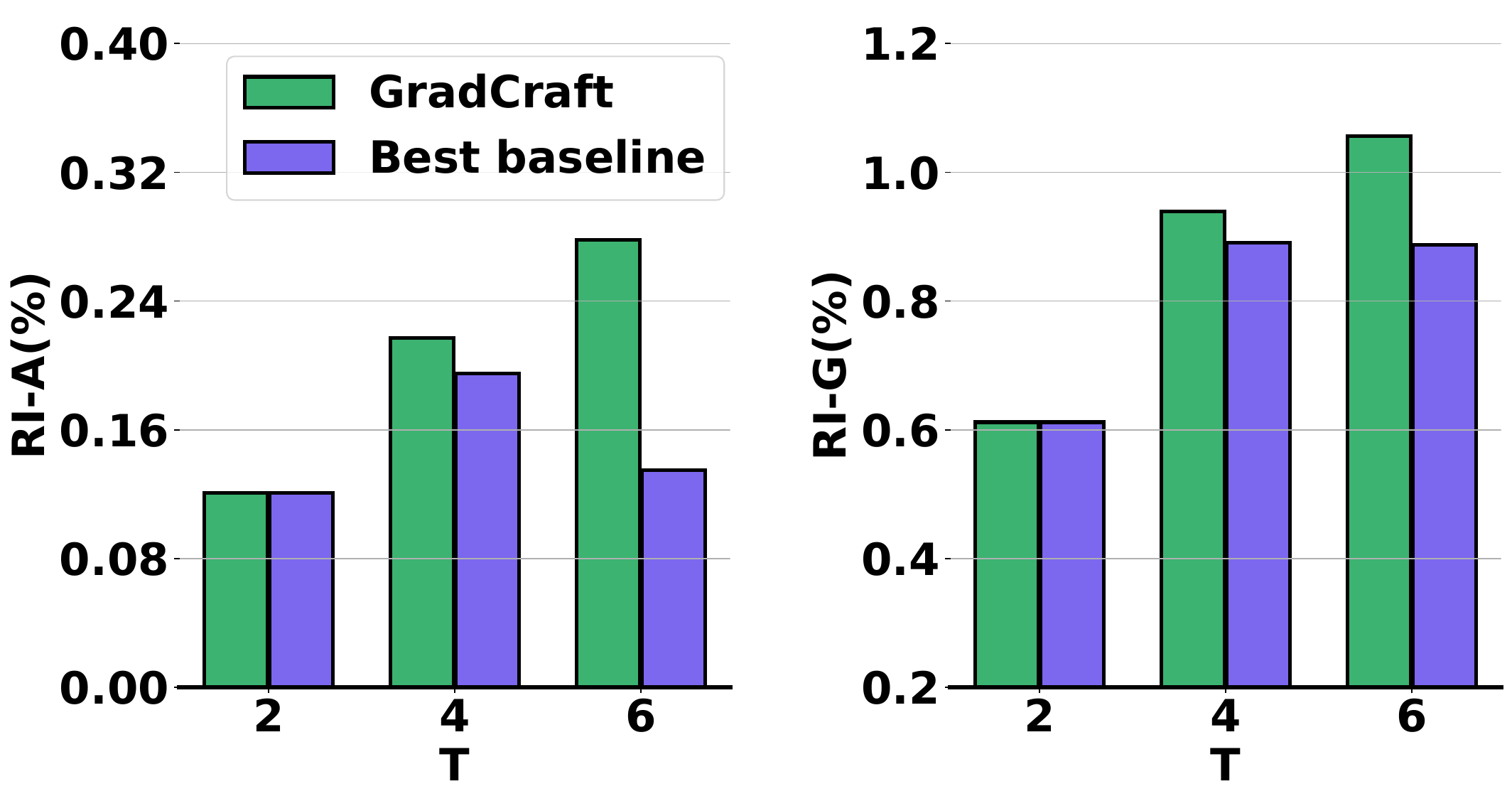}
    \caption{Results of the performance of GradCraft in comparison with the best baseline across different task number $T$ on Wechat.}
    \label{fig:hyper_T}
\end{figure}

In multi-task recommendations, the degree of gradient imbalance is intricately linked to the task number, with higher task numbers leading to an increase in the number of potential conflicting task pairs. Consequently, we conduct a comprehensive study to evaluate the impact of varying task numbers on GradCraft's performance. We also present the performance of the best baseline for comparative analysis. Specifically, we adjust the task number in the range of $\{2, 4, 6\}$ while ensuring an equal number of viewing labels and engagement labels. For $T=2$, we designate EV and Like as the tasks, and for $T=4$, we incorporate EV, LV, Like, and Follow. For $T=6$, we use all the labels mentioned. We depict the relative improvement metrics RI-A and RI-G in Figure~\ref{fig:hyper_T}, and our observations are as follows:

\begin{itemize}[leftmargin=*]
    \item Both the metrics of GradCraft exhibit a consistent increase with the task number. In contrast, the best baseline method does not display a similar trend. This stark contrast suggests that GradCraft possesses a unique capability to achieve gradient balance, which scales up effectively with the increasing complexity introduced by a growing number of tasks. Consequently, GradCraft showcases its potential for practical application in complex recommendation scenarios. This enhanced performance can be attributed to the implementation of flexible magnitude adjustment and thorough direction conflict elimination in GradCraft.
    
    \item Moreover, as the number of tasks increases, the performance gap between GradCraft and the best baseline method widens. This expanding gap provides further evidence supporting the advantages of GradCraft in achieving both appropriate magnitude balance and global direction balance at the gradient level. It is worth mentioning that for $T=2$, both methods yield similar results in terms of the RI-A and RI-G metrics. This similarity can be attributed to the fact that when there is only one pair of tasks, the global projection method employed by GradCraft closely resembles the normal conflict projection method. This consistency aligns with the earlier discussion presented in Equation~\eqref{eq:pcgrad}.
\end{itemize}

\subsection{Online Experiment (RQ5)}

We conduct an online A/B experiment on our production platform, leveraging traffic from over 15 million users. We assesse three key business and engagement metrics: the average time users spend watching videos (WT), the number of effective video viewing records (VV), and the instances of video sharing (Share). Our findings, presented in Table~\ref{exp:online}, demonstrate notable performance enhancements achieved by our method compared to the state-of-the-art multi-task learning baseline implemented in Kuaishou. 

\begin{table}[t]
\caption{Results of the online experiment conducted over one week. It is noteworthy that performance improvements exceeding 0.1\% for WT and VV, and 1.0\% for Share, are considered significant~\cite{PEPNet}.}
\label{exp:online}
\begin{tabular}{cccc}
\hline
          & WT       & VV       & Share    \\ \hline
Base      & -        & -        & -        \\
GradCraft & \textbf{+0.505\%} & \textbf{+0.950\%} & \textbf{+1.746\%} \\ \hline
\end{tabular}
\end{table}
\vspace{-5pt}
\section{Related Work}
In this section,  we navigate through existing research on multi-task learning, encompassing the optimization methodologies and model architectures. Our particular emphasis is on their application within the realm of recommender system.

\subsection{Multi-task Optimization}

Multi-task learning necessitates the simultaneous optimization of multiple tasks. Prior research has proposed various optimization methods to mitigate the imbalance among different tasks, broadly categorized into two lines. The first category involves reweighting loss, adjusting the gradient magnitudes based on different aspects of the specific criteria~\cite{survey,GradNorm,revisiting,Uncertainty,DWA,DBMTL}. For example, UC~\cite{Uncertainty} adjusts the loss weights according to the uncertainty associated with each task, while DWA~\cite{DWA} adapts the loss weights by taking into account the rate of change of the loss value. The second category focuses on manipulating gradient directions to diminish the direction conflict~\cite{MGDA,PCGrad,CAGrad,IMTL,GradVac}. For instance, MGDA~\cite{MGDA} manipulates gradients to achieve a local Pareto optimal solution. PCGrad~\cite{PCGrad} addresses gradient interference by pair-wise projections. CAGrad~\cite{CAGrad} identifies the optimal update vector within a sphere around the average gradient and maximizes the worst local improvement between tasks. IMTL~\cite{IMTL} learns weights to ensure that the aggregated gradient has equal projections onto each task gradient. Among the mentioned works, CAGrad implicitly considers the gradient magnitude. However, it only imposes restrictions on the magnitude of the update vector, rather than finely modifying the magnitudes of each individual task like our proposed GradCraft.

In recent times, there has been a growing focus on developing tailored strategies specifically for the recommender system~\cite{PE-LTR, MetaBalance, SoFA, LabelCraft}, with a particular emphasis on diverse optimization objectives. PE-LTR~\cite{PE-LTR} introduces a Pareto-efficient algorithmic framework for e-commerce recommendations. LabelCraft~\cite{LabelCraft} proposes a labeling model that aligns with the objectives of short video platforms. MetaBalance~\cite{MetaBalance} aims to achieve equilibrium among auxiliary losses by manipulating their gradients to enhance knowledge transfer for the target task. SoFA~\cite{SoFA} optimizes item-side group fairness while maintaining recommendation accuracy constraints. Among these works, MetaBalance bears resemblance to our GradCraft as it incorporates adjustments to gradient magnitudes. However, MetaBalance primarily focuses on multi-behavior learning and solely optimizes performance on the target task. Additionally, it rigidly employs the gradients of the target task as adjustment criteria. In contrast, GradCraft focuses on the optimization of multiple objectives and dynamically utilizes the maximum norm of gradients across all tasks, resulting in greater adaptability and versatility.

\subsection{Multi-task Model}

Multi-task models aim to excel in multiple interrelated tasks simultaneously, extracting shared information to enhance proficiency in each task. While hard parameter sharing models are commonly used, they may suffer from detrimental transfer effects due to task disparities. To address this, soft parameter sharing models have been introduced, such as the cross-stitch network~\cite{cross-stitch} and sluice network~\cite{sluice}, which combine task-specific hidden layers using linear combinations. Gating and attention mechanisms have also been utilized for effective information fusion. Examples include MoE~\cite{MoE}, which uses a gate structure to combine various experts, and MTAN~\cite{DWA}, which incorporates task-specific attention modules within a shared network.

In recommendations, hard parameter sharing at the bottom (SharedBottom)~\cite{DTRN} remains pervasive owing to its simplicity and efficiency, effectively addressing the oversight of task correlations in traditional models rooted in collaborative filtering and matrix factorization~\cite{NCF,NGCF,LightGCN,IFRU}. MMoE~\cite{MMoE} goes a step further by sharing all experts across diverse tasks, utilizing distinct gates for each task to augment the capabilities of the MoE framework. Conversely, ESMM~\cite{ESMM} adopts a soft parameter sharing structure, simultaneously optimizing two correlated tasks through sequential modes to mitigate the sparsity inherent in the prediction target. Expanding upon the shared experts paradigm in MMoE, PLE~\cite{PLE} establishes independent experts for each task, and adopts multi-level extraction networks with progressive separation routing. Furthermore, AdaTT~\cite{AdaTT} enhances its capability by utilizing an adaptive fusion mechanism, enabling the model to more effectively select fine-grained feature representations for individual tasks. Our work diverges from the aforementioned research as it concentrates on optimization perspective and remains model-agnostic.

\section{Conclusion}

This study investigated the application of multi-task learning methods in the recommender system. Recognizing the distinct characteristics of recommendations, we proposed GradCraft to simultaneously achieve an appropriate magnitude balance and a global direction balance to enhance the multi-task optimization. GradCraft dynamically adjusted the gradient magnitudes to align with the maximum gradient norm to establish the appropriate magnitude balance, mitigating interference from gradient magnitudes for subsequent manipulation. Subsequently, it employed projections to eliminate gradient conflicts in directions while considering all conflicting tasks concurrently, thereby ensuring global direction balance. Extensive experiments conducted on both real-world datasets and our production platform provided empirical evidence of its effectiveness in enhancing multi-task recommendations.

In our future work, we will enhance the comprehensiveness of our method by integrating the resolution of conflicting gradients with the improvement of consistency among other gradients. Additionally, we plan to apply our method to other domains, including Computer Vision (CV)~\cite{CV} and Natural Language Processing (NLP)~\cite{NLP}, in order to evaluate its general applicability. Moreover, we recognize the complexity of industrial recommendation scenarios and will focus on developing more effective multi-task learning methods tailored for large-scale industrial settings.

\begin{acks}
    This work is supported by the National Key Research and Development Program of China (2022YFB3104701), the National Natural Science Foundation of China (62272437), and the CCCD Key Lab of Ministry of Culture and Tourism.
\end{acks}

\bibliographystyle{ACM-Reference-Format}
\balance
\bibliography{8_reference}


\end{document}